\documentclass[12pt]{article}
\usepackage[dvips]{color}
\usepackage{amsmath}
\usepackage{graphicx}
\usepackage{subfigure}
\usepackage{epsfig}
\usepackage{amsmath}
\usepackage{cite}
\usepackage{color}
\usepackage{subfigure}

\begin{document}
\begin{center}
\Large{\bf Anisotropic constant-roll inflation with noncommutative model and swampland conjectures}\\
\small \vspace{1cm} {\bf J. Sadeghi$^{\star}$\footnote {Email:~~~pouriya@ipm.ir}}, \quad
{\bf S. Noori Gashti$^{\star}$\footnote {Email:~~~saeed.noorigashti@stu.umz.ac.ir}}, \quad
\\
\vspace{0.5cm}$^{\star}${Department of Physics, Faculty of Basic
Sciences,\\
University of Mazandaran
P. O. Box 47416-95447, Babolsar, Iran}\\
\small \vspace{1cm}
\end{center}
\begin{abstract}
In this paper, we study a constant-roll inflationary model in the presence of a noncommutative  parameter with a homogeneous scalar field minimally coupled to gravity. The specific noncommutative inflation conditions proposed new consequences. On the other hand, we use anisotropic conditions and find new anisotropic constant-roll solutions with respect to noncommutative parameter. Also, we will plot some figures with respect to the specific values of the corresponding parameter and the swampland criteria which is raised from the exact potential obtained from the constant-roll condition. Finally, different of figures lead us to analyze the corresponding results and also show the effect of above mentioned parameter on the inflationary model.
\end{abstract}
\newpage

\section{Introduction}
As we know, Einstein's theory of gravity is one of the most important theories that led to a major breakthrough in particle physics and cosmology. From high energy point of view such theory is not suitable candidate for the describing universe. In that case, the researcher try to modified the theory of gravity in different approaches. On of the important approach proposed by Snyder\cite{1,2}, which  contains attractive properties and leads to modifications the renormalizability of the properties of relative quantum field theory\cite{3,4,5,6,7,8,9,10,11,12,13,14}. Such approach related to noncommutative (NC) of space time coordinates \cite{1,2}. In the Snyder method, the NC parameter  lead us to the short length cutoff. Given the above explanation and concepts about NC , we can say that the gravity will be very important in the scale of NC parameter. Gravity and its effects  with the corresponding length have recently been considered by researchers. Also the deformed phase space scenarios can also be used to study the various features of universe. Therefore, we should also expect such corrections  can be related to space-time in relation to effects of CMB spectrum\cite{15,16,17,18,19,20,21,22}. So, in future  these effects may be seen at cosmological observational data. The use of NC on inflation,  power law inflation and CMB has been discussed by several researchers which can be referred to\cite{23,24,25,26,27,28}. We note here, the subject of inflation is studied from different points of view and different types of models. Also it  has been considered from different perspective such as slow-roll, ultra-slow-roll and constant roll. Recently, constant-roll inflation is determined by constant-roll conditions, ie $\ddot{\phi}=\beta H \dot{\phi}$ which $\beta$ is a constant parameter. Under constant-roll conditions, the equations are obtained with complete accuracy. Hence, many constant-roll inflation models can be consistent with observable data\cite{29,30,31}. Recently, a lot of work has been done in this field. For example, the idea of constant-roll inflation can be extended to the teleparallel $ f (t) $ gravity \cite{32,33,34,35}. Other types of inflation models and effective theories such as $F(T)$ and $F(Q)$ have been recently studied and its cosmological implications have been carefully investigated. On the other hand, some researchers employed the weak gravity conjecture for the investigation of  different attributes of cosmology as inflation models, dark energy and the other implications. The  weak gravity conjecture and the swampland program has been studied in many cosmological concepts, we see \cite{36,37,38,39,40,41,42,43,44,45,46,47,48,49,50,51,52,53}.
Given the above concepts, we want to investigate a constant-roll inflation model of  homogeneous scalar field minimally coupled to gravity with NC . On the other hand, the study of a series of inflation models with form of anisotropy and will lead to the expansion of the universe. Such inflation models that are examined in supergravity the content which are called anisotropic inflation\cite{54}. Various models of this type of inflation have been reviewed in literature\cite{55,56,57,58,59,60,61,62,63,64,65,66,67,68,69,70,71,72,73,74,75,76,77,78,79,80,81,82,83,84,85,86,87,88,89,90,91,92,93,94,95,96,97,98}. In this paper, we examine inflation model with constant-roll conditions. By obtaining the exact solutions, we  analyze the swampland criteria according to different values of NC parameter. Finally, we plot different figures according to the different values of the NC parameter. And also, we show the effect of such parameter on the corresponding model. Also, the relation of NC parameter and swampland condition lead us to consider some special values for the corresponding parameter $\theta.$

\section{Non-commutativity of inflation model}
First, we briefly introduce the inflation model and  apply the NC to the particular type of inflation model. Applying this form and other NC frameworks, both at the classical and quantum level in cosmological settings, provides a rational for studding very important challenges in cosmology, such as the study of UV / IR mixing. It was called a means of describing physical phenomena at different distances, or  in different energy regimes. This can be deduced very nice result from the use of NC quantum field theories\cite{99,100,101,102,103,104,105,106,107,108}. Therefore, with respect to the characteristics of NC in cosmology can be a serious motivate to study many unknown phenomena and discover effective and promising solutions. In the present work we actually limit to the classical geometric framework, where the corresponding NC effects are obtained using conventional features and using Poisson brackets. Recently, different applications of NC  on cosmology have been investigated  and for the study of  gravitational collapse of a homogeneous scalar field have been  obtained elegance results\cite{18,20}. Before we are going to NC cosmological scenario, we first write the Lagrangian of scalar field $\phi$ minimally coupled to gravity, so the corresponding Lagrangian is,
\begin{equation}\label{1m}
l=\sqrt{-g}\left[\frac{1}{16\pi G} R-\frac{1}{2}g^{\mu\nu}\phi_{,\mu}\phi_{,\nu}-V(\phi)\right]
\end{equation}
In order to apply the NC to the above system, we prefer to write the Hamiltonian of the corresponding system. Here, we need to introduce background space-time which is described by the following spatially flat FLRW universe,
\begin{equation}\label{1n}
ds^{2}=-N^{2}dt^{2}+a^{2}(t)\left(dx^2+dy^2+dz^2\right)
\end{equation}
where $t$ is cosmic time, $x$, $y$, $z$ are the Cartesian coordinates, $N(t)$ is a lapse function an $a(t)$ is the scale factor.
Now we use equation (2) and obtain the $R$, $g$ and $g^{\nu\mu}$, finally the known Lagrangian density of (1) becomes,
\begin{equation}\label{1o}
l=-\frac{3}{8\pi G}N^{-1}a{\dot{a}}^2+\frac{1}{2}N^{-1}a^3 {\dot{\phi}}^2-Na^3 V(\phi)
\end{equation}
So, the equation (3) lead us to obtain the corresponding  Hamiltonian of this model, which is given by \cite{20}
\begin{equation}\label{1}
\textbf{H}=-\frac{2}{3}\pi GNa^{-1}P_{a}^{2}+\frac{1}{2}Na^{-3}P^{2}_{\phi}+Na^{3}V(\phi)
\end{equation}
where $\textbf{H}$, $P_{a}$ and $P_{\phi}$ are Hamiltonian,  momenta conjugates of scale factor and  momenta conjugates of scalar field respectively. Also, here we take the comoving gauge where $N-1$. The equation of motion correspond to the phase space coordinates $\{a, \phi, p_{a}, p_{\phi}\}$with respect to equation (1) we will have ,

\begin{equation}\label{2}
\begin{split}
&\dot{a}=\{a,\textbf{H}\}=-\frac{4\pi G}{3}a^{-1}P_{a}\\
&\dot{P}_{a}=\{P_{a},\textbf{H}\}=-\frac{2G\pi}{3}a^{-2}P_{a}^{2}+\frac{3}{2}a^{-4}P_{\phi}^{2}-3a^{-2}P_{a}^{2}V(\phi)\\
&\dot{\phi}=\{\phi,\textbf{H}\}=a^{-3}P_{\phi}\\
&\dot{P}_{\phi}=\{P_{\phi},\textbf{H}\}=-a^{3}V'(\phi)\\
\end{split}
\end{equation}
In order to study the cosmological stuff we have to write the following equations which is obtained by the above information,
\begin{equation}\label{71}
H^{2}=\frac{8\pi G}{3}\left[\frac{1}{2}\dot{\phi}^{2}+V(\phi)\right]\equiv\frac{8\pi G}{3}\rho_{tot},
\end{equation}

\begin{equation}\label{81}
2\frac{\ddot{a}}{a}+H^{2}=-8\pi G\left[\frac{1}{2}\dot{\phi}^{2}-V(\phi)\right]\equiv-8\pi G p_{tot},
\end{equation}
and
\begin{equation}\label{91}
\ddot{\phi}+3H\dot{\phi}+V'(\phi)=0.
\end{equation}
where $H$ is Hubble parameter.

Now we consider special of canonical non-commutativity and concentrate to the suitable deformation on classical phase space. This lead us to write the following  equation,
\begin{equation}\label{2m}
\{P_{a}, P_{\phi}\}=\theta \phi^3
\end{equation}
where $\theta$ is NC parameter and it has been assumed as a constant. Here, we note that if we take non-commutativity on the momenta, we will have modified scalar field and scale factor. But, if we take the non-commutativity to the field and scale factor, we will see any NC effect will be absent for a vanishing potential. The dynamical non-commutativity between momenta more interesting than the choice of first one. It means that when we take non-commutativity between momenta the effect of $\theta$ parameter more than case of we take non-commutativity on the field and scale factor. So, all above explantation give us opportunity to take equation following  modified  equations  of motion\cite{20},
\begin{equation}\label{3}
\dot{P}_{a}=-\frac{2\pi G}{3}a^{-2}P_{a}^{2}+\frac{3}{2}a^{-4}P_{\phi}^{2}-3a^{2}V(\phi)+\theta(a^{-3}\phi^{3}P_{\phi}),
\end{equation}
and
\begin{equation}\label{4}
\dot{P}_{\phi}=-{a}^{3}V'(\phi)+\theta(\frac{4\pi G}{3}{a}^{-1}\phi^{3}P_{a}).
\end{equation}
As we see the equations (2) are not modified by considering any non-commutativity. But the modified equation (7) and (8) obtained by the following auxiliary formulas and equation (6),
\begin{equation}\label{5}
\{P_{a},f(P_{a},P_{\phi})\}=\theta\phi^{3}\frac{\partial f}{\partial P_{\phi}}
\end{equation}
and
\begin{equation}\label{6}
\{P_{\phi},f(P_{a},P_{\phi})\}=-\theta\phi^{3}\frac{\partial f}{\partial P_{a}}
\end{equation}

In case of  $\theta\rightarrow 0$,  we have  standard commutative equations of motion. So with respect to above concept, the equations of motion at NC framework  can obtained by the following equations,
\begin{equation}\label{7}
H^{2}=\frac{8\pi G}{3}\left[\frac{1}{2}\dot{\phi}^{2}+V(\phi)\right]\equiv\frac{8\pi G}{3}\rho_{tot},
\end{equation}

\begin{equation}\label{8}
2\frac{\ddot{a}}{a}+H^{2}=-8\pi G\left[\frac{1}{2}\dot{\phi}^{2}-V(\phi)+\frac{\theta\dot{\phi}\phi^{3}}{3a^{2}}\right]\equiv-8\pi G p_{tot},
\end{equation}
and
\begin{equation}\label{9}
\ddot{\phi}+3H\dot{\phi}+V'(\phi)+\theta H(\frac{\phi^{3}}{a^{2}})=0
\end{equation}
where we  consider $8\pi G=1$. The energy density and pressure with respect to the scalar field for NC model are determined by $\rho_{tot}$ and $p_{tot}$. In that case the equation state depend to NC parameter $\theta$ and we just keep $\theta$ as lower order. Because such parameter is very short length with respect to any parameter in here. Also we note that in  case of  $\theta=0$ we have usual cosmological equations as (6), (7) and (8).
\section{Anisotropic constant-roll inflation}
In this section, we will try to obtain exact solutions for the  Hubble parameter,  potential and the other important parameters which play an important roll for the explantation of inflation. Here, we study the constant-roll inflationary models in the presence of non-commutative geometry. So, we expect anisotropy of the expansion of universe due to existence of non-commutative space. As we  know,  the anisotropy in context of cosmology including inflationary model discussed from supergravity point of  view. On the other hand,  we shall make two crucial assumptions to have constant-roll conditions. First, we have to consider the first slow-roll index is very small during the inflation era. In the second step, we shall assume that although the first slow-roll index $\epsilon_{1} \ll 1$, in that case the constant-roll condition holds true. So with respect to equation (14) and (15), one can obtain,

\begin{equation}\label{10}
2\frac{\ddot{a}}{a}-2H^{2}=-\dot{\phi}^{2}-\frac{\theta\phi^{3}\dot{\phi}}{3a^{2}}
\end{equation}

According to the definition of the Hubble parameter that means $H\equiv \frac{\dot{a}}{a}$ and using the above equation we will have,

\begin{equation}\label{11}
2\dot{H}=-\dot{\phi}^{2}-\frac{\theta\phi^{3}\dot{\phi}}{3a^{2}}
\end{equation}

Now, we use  $\dot{H}=\frac{d H}{d \phi}\dot{\phi}$ and put to equation (18), one can rewrite following equation,
\begin{equation}\label{12}
\dot{\phi}^{2}+2\frac{d H}{d \phi}\dot{\phi}+\frac{\theta\phi^{3}\dot{\phi}}{3a^{2}}=0
\end{equation}

We take derivative of $\phi$ with respect to time, so we have,
\begin{equation}\label{13}
\ddot{\phi}=-2\frac{d^{2}H}{d\phi^{2}}\dot{\phi}-\frac{\theta\phi^{2}\dot{\phi}}{a^{2}}
\end{equation}
Here, we take advantage from constant-roll condition as $\ddot{\phi}=-(3+\alpha)H \dot{\phi}$ and put to  equation (20), generally we will arrive at,
\begin{equation}\label{14}
-(3+\alpha)H=-2\frac{d^{2}H}{d\phi^{2}}-\frac{\theta\phi^{2}}{a^{2}}
\end{equation}
Here, we assume $\theta=0$, the equation (21) reduce to,
\begin{equation}\label{15}
\frac{d^{2}H}{d\phi^{2}}-\frac{3+\alpha}{2}H=0
\end{equation}
So by solving the equation (22), one can obtain,
\begin{equation}\label{16}
H=C_{1}\exp(\sqrt{\frac{3+\alpha}{2}}\phi)+C_{2}\exp(-\sqrt{\frac{3+\alpha}{2}}\phi)
\end{equation}

The  equation (23) is special solution of $H$, because the effect of non-commutative parameter is cancel. So, according to special solution of (23), we give ansatz for the general solution of equation (21) with effect of non-commutative parameter on the system. So, the general ansatz for the solution of anisotropy inflation will be as,
\begin{equation}\label{17}
H=C_{1}\exp(\lambda(a,\theta)\sqrt{\frac{3+\alpha}{2}}\phi)+C_{2}\exp(-\lambda(a,\theta)\sqrt{\frac{3+\alpha}{2}}\phi)
\end{equation}
By substituting this ansatz in equations of  (19),  (21) and performing a series of straightforward calculations, we obtain the parameter ($\lambda$) and then the Hubble parameter for the NC model. Therefore, we consider a series of special examples according to the boundary conditions and  calculate the mentioned above values. Then we use the Hubble parameter and calculate the exact solution for the potential and other parameters of cosmological model.
\subsubsection*{Case 1}
We consider a special cases of Hubble parameter as $C_{1}=C_{2}=\frac{M}{2}$ which placed in equation (24). This gives two different values for $\lambda$, hence two Hubble parameter is given by,
\begin{equation}\label{18}
H_{1,2}=M\cosh(\frac{1}{2}\phi\sqrt{3+\alpha+\frac{2\mp\frac{\sqrt{8\phi^{4}\theta+a^{2}M(-2+\phi^{2}(3+\alpha))^{2}}}{a\sqrt{M}}}{\phi^{2}}})
\end{equation}
So the above concept of Hubble parameter lead us to have  potential $V(\phi)$ as,
\begin{equation}\label{19}
\begin{split}
&x_{1}=3M(1+\cosh(\frac{1}{2}\phi\sqrt{3+\alpha+\frac{2\mp\frac{\sqrt{8\phi^{4}\theta+a^{2}M(-2+\phi^{2}(3+\alpha))^{2}}}{a\sqrt{M}}}{\phi^{2}}}))\\
&x_{2}=(8\phi^{2}\theta+a^{2}M(3+\alpha)(-2+\phi^{2}(3+\alpha))-a\sqrt{M}(3+\alpha)\sqrt{8\phi^{4}\theta+a^{2}M(-2+\phi^{2}(3+\alpha))^{2}})^{2}\\
&x_{3}=(\sinh(\frac{1}{2}\phi\sqrt{3+\alpha+\frac{2\mp\frac{\sqrt{8\phi^{4}\theta+a^{2}M(-2+\phi^{2}(3+\alpha))^{2}}}{a\sqrt{M}}}{\phi^{2}}}))^{2}\\
&x_{4}=a^{2}(8\phi^{4}\theta+a^{2}M(-2+\phi^{2}(3+\alpha))^{2})(3+\alpha+\frac{2\mp\frac{\sqrt{8\phi^{4}\theta+a^{2}M(-2+\phi^{2}(3+\alpha))^{2}}}{a\sqrt{M}}}{\phi^{2}})\\
&V_{1,2}(\phi)=\frac{1}{2}M(x_{1}-\frac{x_{2}\times x_{3}}{x_{4}})
\end{split}
\end{equation}
also with respect to two equations (24) and (18), one can obtain $\dot{\phi}$, as follows
\begin{equation}\label{20}
\begin{split}
&x_{1}=3a\sqrt{M}(8\phi^{2}\theta+a^{2}M(3+\alpha)(-2+\phi^{2}(3+\alpha))-a\sqrt{M}(3+\alpha)\sqrt{8\phi^{4}\theta+a^{2}M(-2+\phi^{2}(3+\alpha))^{2}})\\
&x_{2}=\sinh(\frac{1}{2}\phi\sqrt{3+\alpha+\frac{2\mp\frac{\sqrt{8\phi^{4}\theta+a^{2}M(-2+\phi^{2}(3+\alpha))^{2}}}{a\sqrt{M}}}{\phi^{2}}})\\
&x_{3}=\sqrt{8\phi^{4}\theta+a^{2}M(-2+\phi^{2}(3+\alpha))^{2}}\times\sqrt{3+\alpha+\frac{2\mp\frac{\sqrt{8\phi^{4}\theta+a^{2}M(-2+\phi^{2}(3+\alpha))^{2}}}{a\sqrt{M}}}{\phi^{2}}}\\
&\dot{\phi}_{1,2}=\frac{-\phi^{3}\theta+\frac{x_{1}\times x_{2}}{x_{3}}}{3a^{2}}
\end{split}
\end{equation}
According to the above equation, we solve the above relation and obtain the $\phi$ in terms of $t$. The scale factor and $\alpha$ will  be easily calculated according to the definition of the Hubble parameter, $H=\frac{\dot{a}}{a}$. Next, we will consider another example.

\subsubsection*{Case 2}
Now we consider $C_{1}=\frac{M}{2},C_{2}=-\frac{M}{2}$, so in this part, two different values for $\lambda$ are obtained. Therefore, the values obtained for other parameters as $H$, $V(\phi)$ and $\dot{\phi}$ have two different values as in the previous part, which are given by,
\begin{equation}\label{21}
H_{1,2}=M\sinh(\frac{1}{2}\mp\frac{\sqrt{\frac{M(3+\alpha)^{2}(4\phi^{2}\theta+a^{2}M(3+\alpha))}{a^{2}}}}{2M(3+\alpha)^{\frac{3}{2}}})
\end{equation}

So with respect to above concept the potential $V(\phi)$ and $\dot{\phi}$ are calculated exactly like the previous part.

\begin{equation}\label{22}
\begin{split}
V_{1,2}(\phi)=&M(-\frac{8\phi^{2}\theta^{2}(\cosh(\frac{1}{2}\mp\frac{\sqrt{\frac{M(3+\alpha)^{2}(4\phi^{2}\theta+a^{2}M(3+\alpha))}{a^{2}}}}{2M(3+\alpha)^{\frac{3}{2}}}))^{2}}{a^{2}(3+\alpha)(4\phi^{2}\theta a^{2}M(3+\alpha))}\\
&+3M(\sinh(\frac{1}{2}\mp\frac{\sqrt{\frac{M(3+\alpha)^{2}(4\phi^{2}\theta+a^{2}M(3+\alpha))}{a^{2}}}}{2M(3+\alpha)^{\frac{3}{2}}}))^{2})
\end{split}
\end{equation}
and
\begin{equation}\label{23}
\dot{\phi}_{1,2}=-\frac{\phi^{3}\theta}{3a^{2}}+\frac{4M\phi\theta\sqrt{3+\alpha}\cosh(\frac{1}{2}\mp\frac{\sqrt{\frac{M(3+\alpha)^{2}(4\phi^{2}\theta+a^{2}M(3+\alpha))}{a^{2}}}}{2M(3+\alpha)^{\frac{3}{2}}})}{a^{2}\sqrt{\frac{M(3+\alpha)^{2}(4\phi^{2}\theta+a^{2}M(3+\alpha))}{a^{2}}}}
\end{equation}
\section{Swampland conjectures and anisotropy inflation model }
In this section, by using exact solutions of potential which is obtained by constant - roll condition in presence of non-commutative space in previous section, we investigate  the swampland $dS$ conditions. Therefore, as stated in the literature, the swampland $dS$ conditions are expressed in the following form\cite{49}.
\begin{equation}\label{24}
M_{pl}\frac{V'}{V}>c_{1}\hspace{2cm} Case(I),
\end{equation}
and
\begin{equation}\label{25}
M_{pl}^{2}\frac{V''}{V}<-c_{2} \hspace{2cm} Case(II)
\end{equation}
where $c_{1}$ and $c_{2}$ are constants of order unity. Now, by considering the swampland conditions as well as the exact solution of the potential from the constant-roll condition, we examine the rate of change these conditions with respect to the different values of the NC parameter. So for the (case I) as well as from equations (26), (31) and (32), we will investigate the changes of the swampland condition according to the NC parameter. In those cases we have some figures.

\begin{figure}[h!]
 \begin{center}
 \subfigure[]{
 \includegraphics[height=6cm,width=7cm]{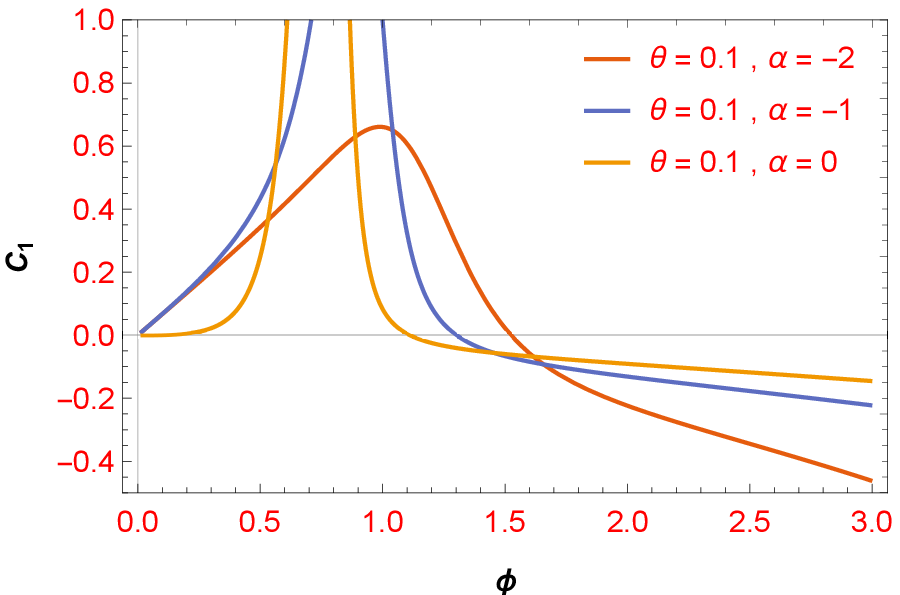}
 \label{1a}}
 \subfigure[]{
 \includegraphics[height=6cm,width=7cm]{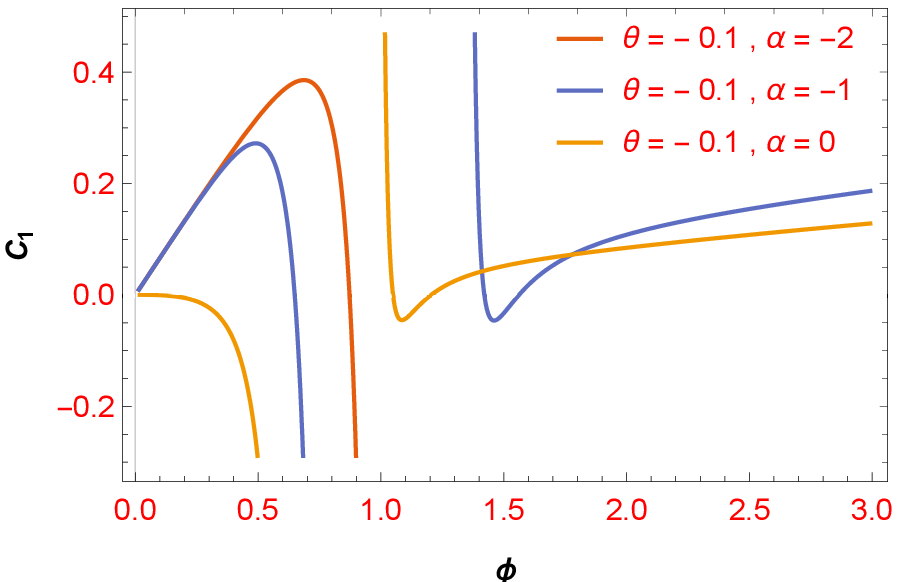}
 \label{1b}}
 \subfigure[]{
 \includegraphics[height=6cm,width=7cm]{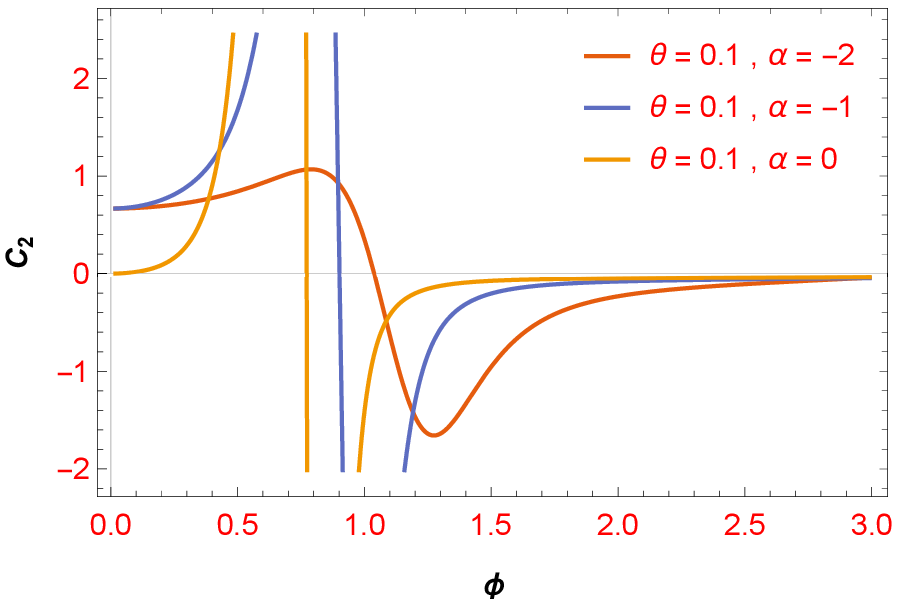}
 \label{1b}}
 \subfigure[]{
 \includegraphics[height=6cm,width=7cm]{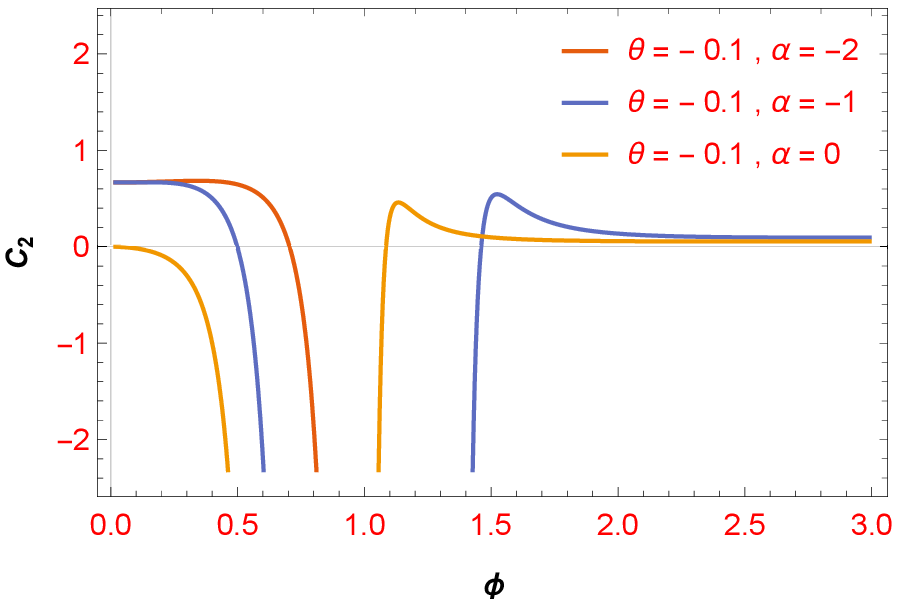}
 \label{1b}}
  \caption{\small{The changes of swampland dS conjecture of $V_{1}$ in term of $\phi$ for case 1 with different values of $\theta$ and $\alpha$}}
 \label{1}
 \end{center}
 \end{figure}

\begin{figure}[h!]
 \begin{center}
 \subfigure[]{
 \includegraphics[height=6cm,width=7cm]{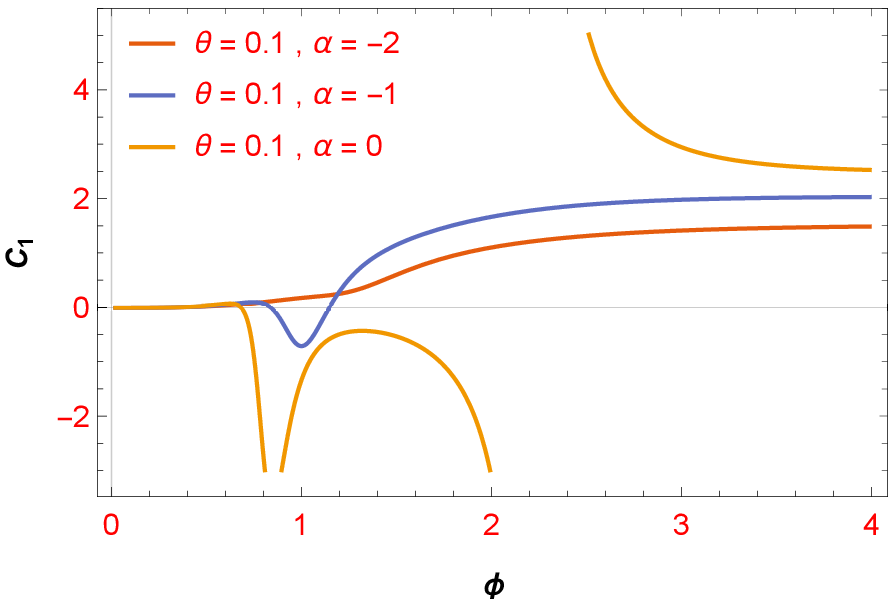}
 \label{1a}}
 \subfigure[]{
 \includegraphics[height=6cm,width=7cm]{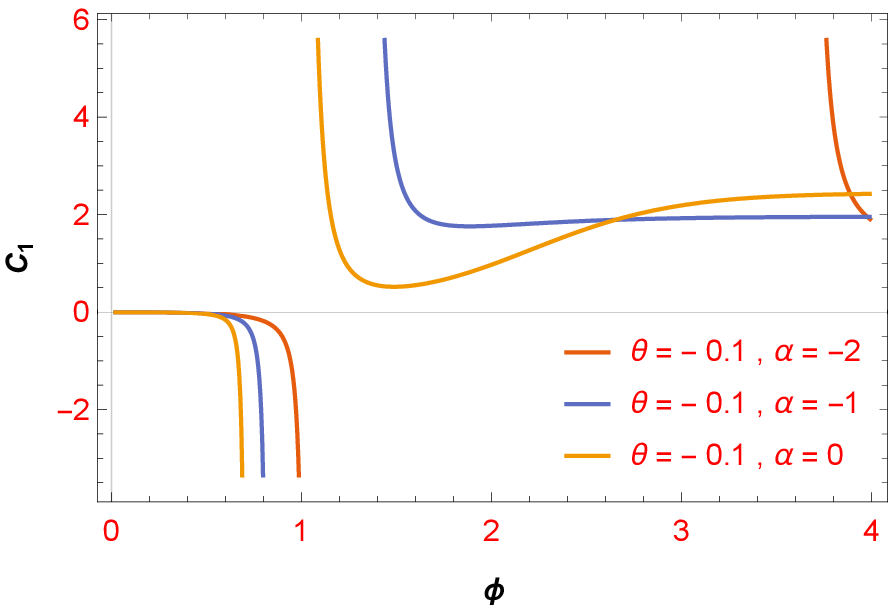}
 \label{1b}}
 \subfigure[]{
 \includegraphics[height=6cm,width=7cm]{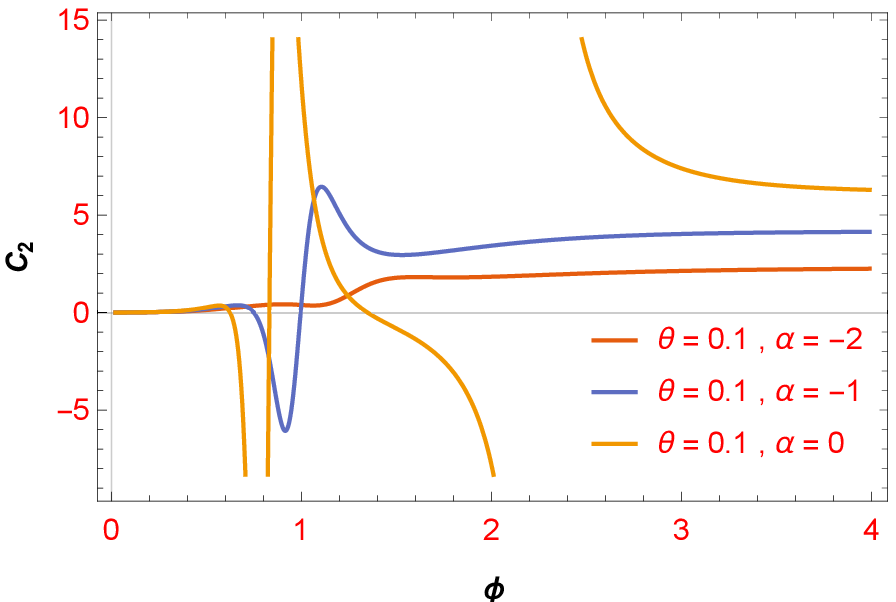}
 \label{1b}}
 \subfigure[]{
 \includegraphics[height=6cm,width=7cm]{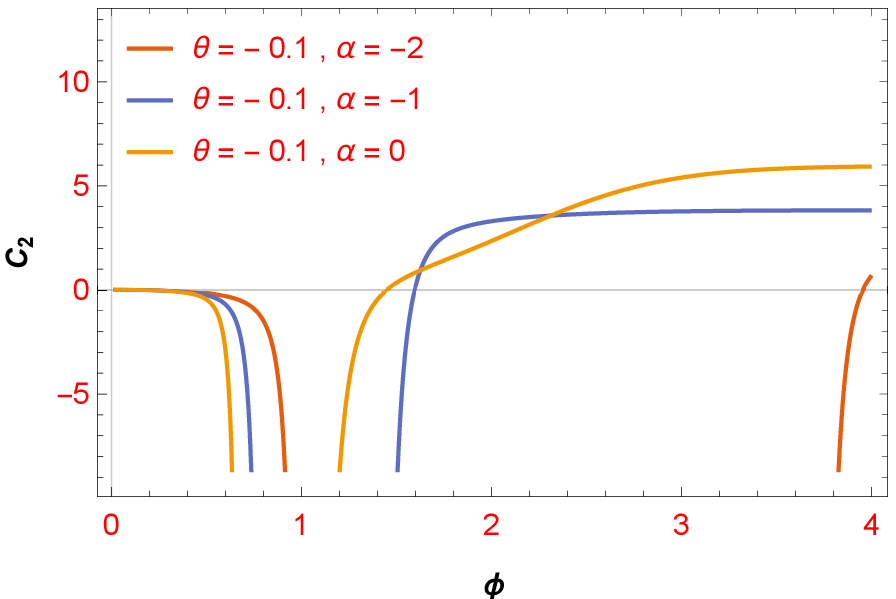}
 \label{1b}}
  \caption{\small{The changes of the swampland dS conjecture of $V_{2}$ in term of $\phi$ for case 1 with different values of $\theta$ and $\alpha$ }}
 \label{2}
 \end{center}
 \end{figure}
\newpage
As we can see in  figures, we investigate the rate of change of each of two dS swampland conditions (I) and (II) with respect to the two different potential values which obtained by the constant-roll condition in  equation (21) with considering different values of NC parameter. On the other hand,  we consider two state. For $\theta>0$, the rate of change  is shown by  figures (a) and (c) and for the negative values of $\theta$, the rate of change  is determined by figures (b) and (d).
We also in figures (1) and (2) show the same change for the potential of (case I) in  equation (26) . In here we also have same situation as a previous case for the $\theta>0$ and  $\theta<0$ which is shown in  figures (3) and (4). We note here,  with respect to equations (29), (31) and (32) one can show,

\begin{figure}[h!]
 \begin{center}
 \subfigure[]{
 \includegraphics[height=6cm,width=7cm]{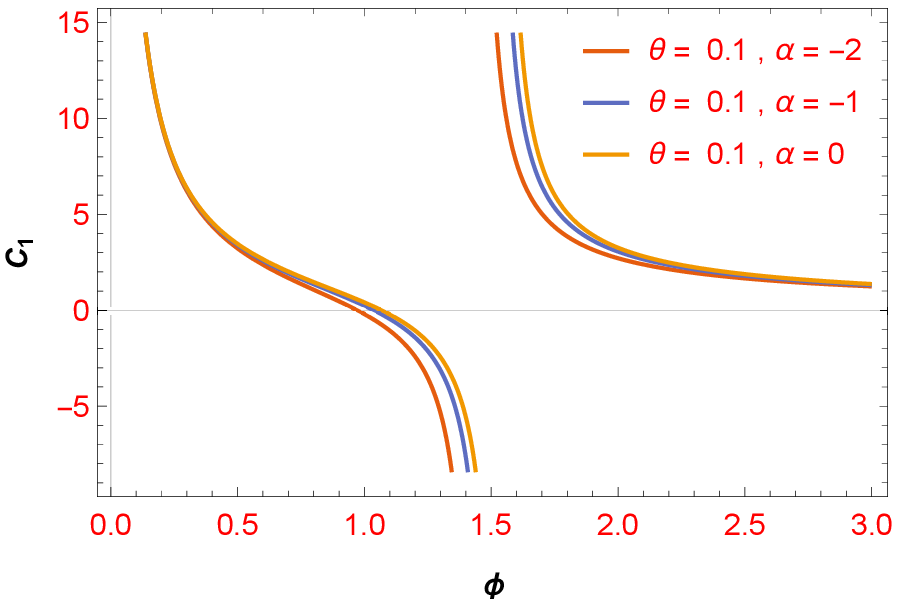}
 \label{1a}}
 \subfigure[]{
 \includegraphics[height=6cm,width=7cm]{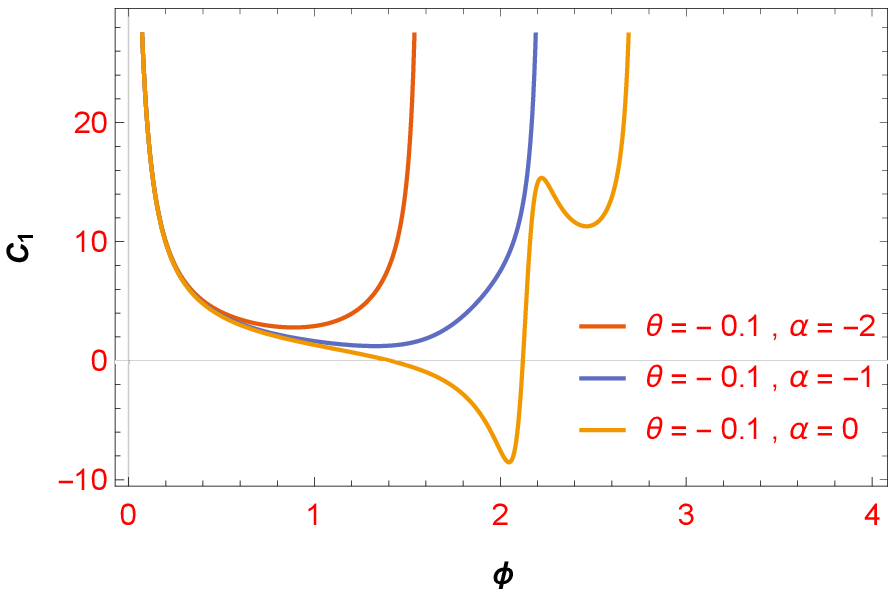}
 \label{1b}}
 \subfigure[]{
 \includegraphics[height=6cm,width=7cm]{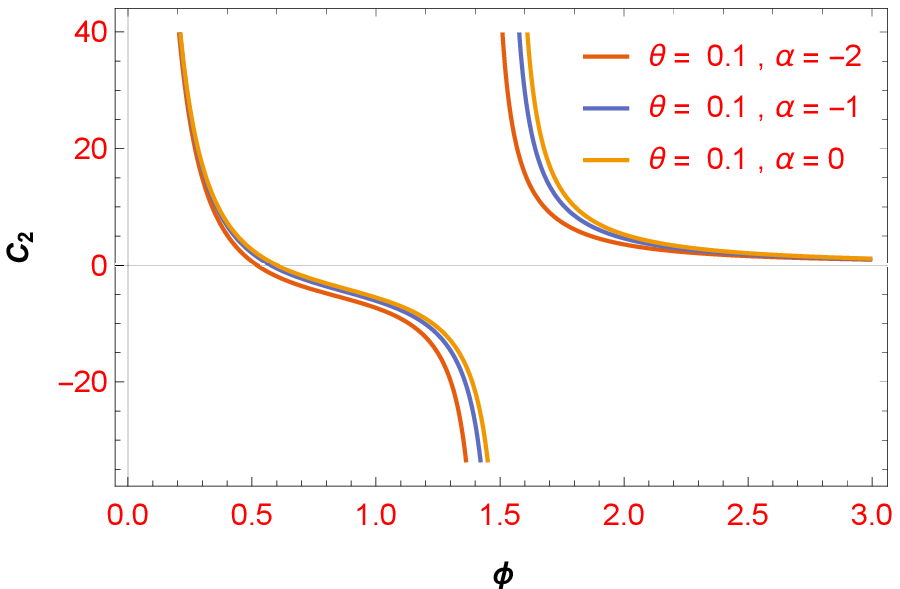}
 \label{1b}}
 \subfigure[]{
 \includegraphics[height=6cm,width=7cm]{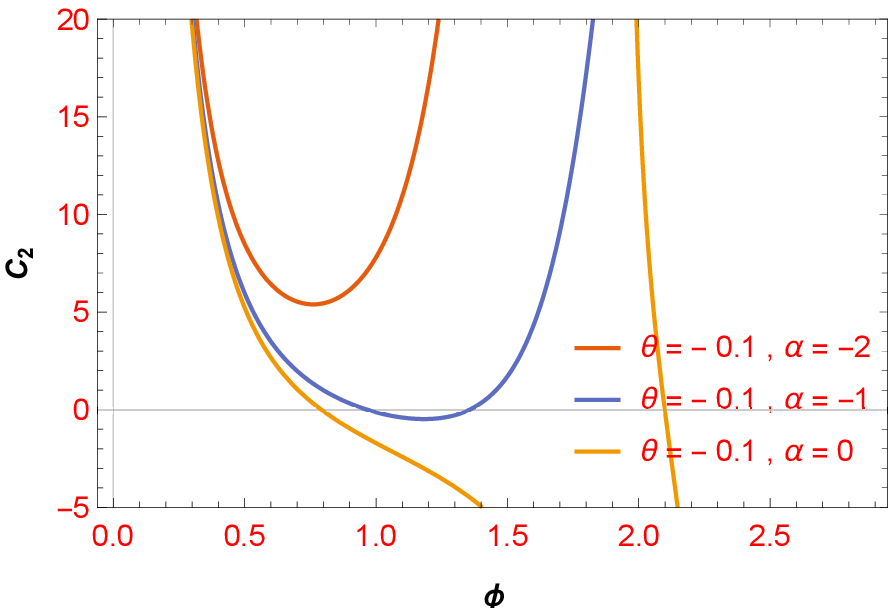}
 \label{1b}}
  \caption{\small{The behaviour of the swampland dS conjecture of $V_{1}$ in term of $\phi$ for case 2 with  different values of $\theta$ and $\alpha$ }}
 \label{3}
 \end{center}
 \end{figure}

\begin{figure}[h!]
 \begin{center}
 \subfigure[]{
 \includegraphics[height=6cm,width=7cm]{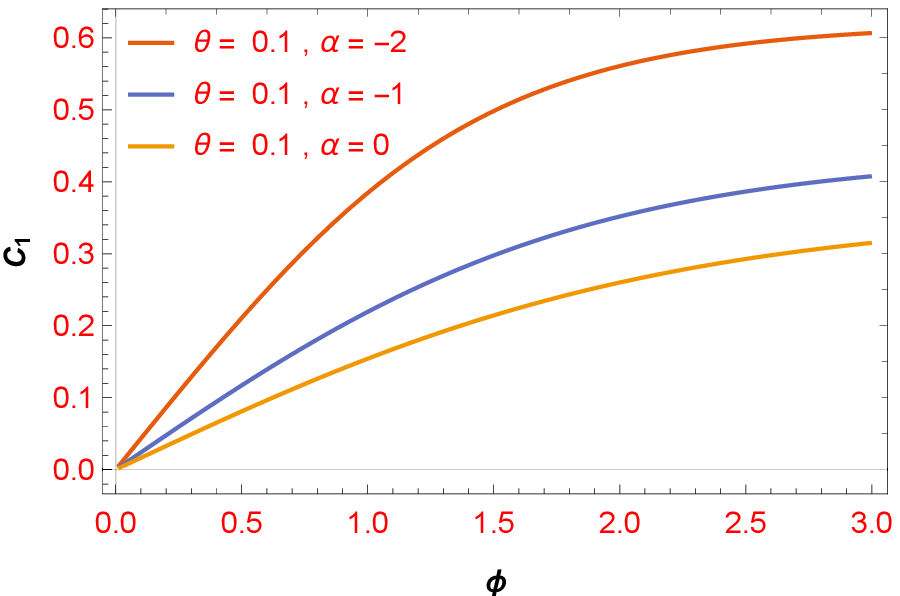}
 \label{1a}}
 \subfigure[]{
 \includegraphics[height=6cm,width=7cm]{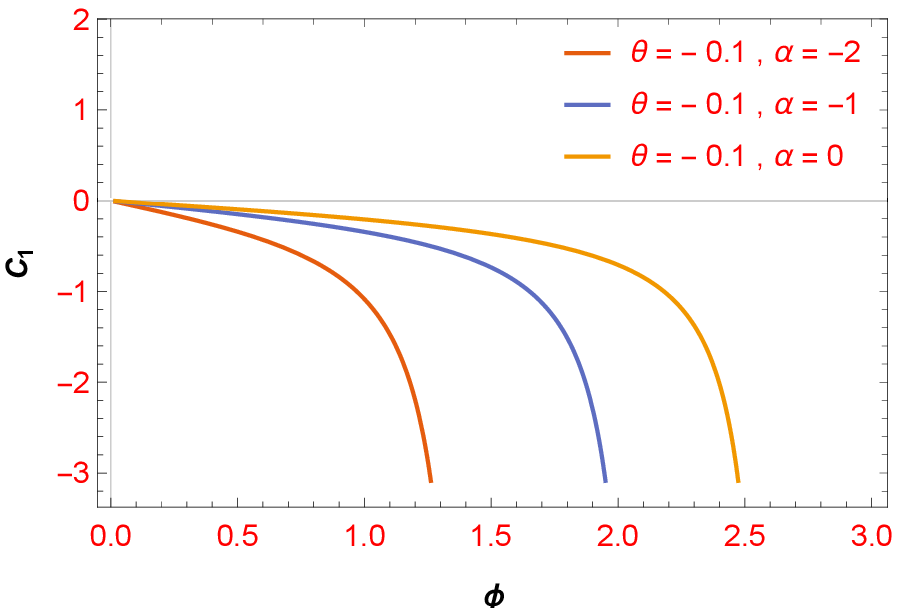}
 \label{1b}}
 \subfigure[]{
 \includegraphics[height=6cm,width=7cm]{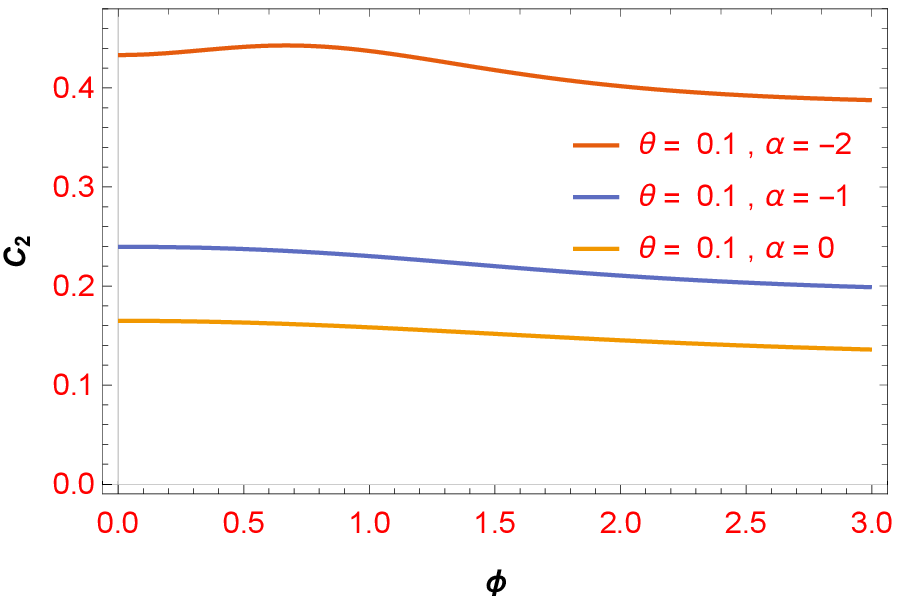}
 \label{1b}}
 \subfigure[]{
 \includegraphics[height=6cm,width=7cm]{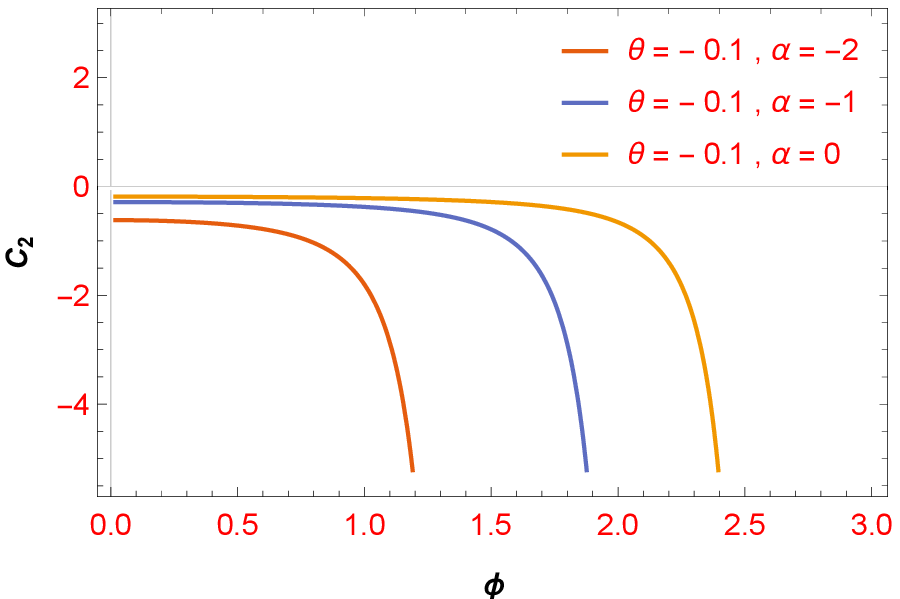}
 \label{1b}}
  \caption{\small{The behaviour of the swampland dS conjecture of $V_{2}$ in term of $\phi$ for case 2 with different values of $\theta$ and $\alpha$ }}
 \label{4}
 \end{center}
 \end{figure}
As you can see in the above figures, the areas consistent with the observable data for the two potentials ($V_{1}$ ,$V_{2}$) has been specified given the acceptable values for $c_{1}$ and $c_{2}$ as well as the values for NC parameter.
According to the above-mentioned issues, it can be noted that leaving the area where the swampland conditions are satisfied coincides by an improvement in the general relativity conditions. It can be very interesting because it makes possible to restore general relativity from theories that Einstein's theory does not work(it means   in strong energy regime). In the other hand, the history of the universe can be traced with curvatures which determined the transition from swampland to general relativity according to parameters such as scalar field or Ricci scalar.
\newpage
\section{Conclusions}

In this paper, we studied a new perspective constant-roll inflation model in the presence of a NC parameter with a homogeneous scalar field minimally coupled to gravity. As we know, the specific NC inflation conditions presented new consequences. On the other hand, we used anisotropic conditions and we found new anisotropic constant-roll solutions with respect to NC parameter. Also, we plotted some figures with respect to the specific values of the corresponding parameter and the swampland criteria which was raised from the exact potential obtained from the constant-roll condition. Finally, we analyzed these results. It will be very interesting to study the scalar and tensor perturbations as well as the role of the NC parameter in calculating cosmological parameters such as tensor-to-scalar ratio, scalar spectrum index,  we leave these calculations to future work.

\end{document}